\documentclass[%
 reprint,
superscriptaddress,
 amsmath,amssymb,
 aps,
]{revtex4-1}

\usepackage{color} 
\usepackage{ulem} 

\usepackage{graphicx}
\usepackage{dcolumn}
\usepackage{bm}

\begin{document}

\preprint{APS/123-QED}

\title{Surface tension and the mechanics of liquid inclusions in compliant solids}

\author{Robert W. Style}
\affiliation{%
Yale University, New Haven, CT 06520, USA 
}
\affiliation{Mathematical Institute, University of Oxford, Oxford, OX1 3LB, UK}

\author{John S. Wettlaufer}%
\affiliation{%
Yale University, New Haven, CT 06520, USA 
}
\affiliation{Mathematical Institute, University of Oxford, Oxford, OX1 3LB, UK}

\author{Eric R. Dufresne}%
\email[]{eric.dufresne@yale.edu}
\affiliation{%
Yale University, New Haven, CT 06520, USA 
}


%

\date{\today}

\begin{abstract}
Eshelby's theory of inclusions has wide-reaching implications across the mechanics of materials and structures including the theories of composites, fracture, and plasticity.
However, it does not include the effects of surface stress, which has recently been shown to control many processes in soft materials such as gels, elastomers and biological tissue.
To extend Eshelby's theory of inclusions to soft materials, we consider liquid inclusions within an isotropic, compressible, linear-elastic solid. 
We solve for the displacement and stress fields around individual stretched inclusions, accounting for the bulk elasticity of the solid and the surface tension (\textit{i.e.} isotropic strain-independent surface stress) of the solid-liquid interface.
Surface tension significantly alters the inclusion's shape and stiffness as well as its near- and far-field stress fields.
These phenomenon depend strongly on the ratio of inclusion radius, $R$, to an elastocapillary length, $L$.  
Surface tension  is significant whenever inclusions are smaller than $100L$.
While Eshelby theory predicts that liquid inclusions generically reduce the stiffness of an elastic solid, our results show that liquid inclusions can actually stiffen a  solid when $R<3L/2$. 
Intriguingly, surface tension cloaks the far-field signature of liquid inclusions when  $R=3L/2$.
These results are have far-reaching applications  from measuring local stresses in biological tissue, to determining the failure strength of soft composites.
\end{abstract}

\pacs{Valid PACS appear here}
\maketitle

\section{Introduction}
Eshelby's theory of inclusions \cite{eshe57} provides a fundamental result underpinning a wide swath of phenomena in composite mechanics \cite{hash63,mori73,hill63,budi65}, fracture mechanics \cite{rice68,budi76}, dislocation theory \cite{mura87}, plasticity \cite{hutc70,berv78} and even seismology \cite{kana75}.
The theory describes how an inclusion of one elastic material deforms when it is embedded in an elastic host matrix.
At an individual inclusion level, it predicts how the inclusion will deform in response to far-field stresses applied to the matrix.
It also allows the prediction of the macroscopic material properties of a composite from a knowledge of its microstructure. 

Eshelby's theory does not include the effects of surface stresses at the inclusion/matrix boundary.
However, recent work has suggested that surface stresses need to be accounted for in soft materials.
This has been suggested both by theoretical models of nanoscale inclusions \cite{shar04,yang04,duan05}, and by 
recent experiments which have shown that surface tension (isotropic, strain-independent surface stress) can also significantly affect soft solids at micron and even millimetric scales.
For example, solid capillarity limits the resolution of lithographic features \cite{hui02,jago12,mora13,pare14}, drives pearling and creasing instabilities  \cite{mora10,mora11,chak13,hena14}, causes the Young-Dupr\'{e} relation to break down for sessile droplets \cite{styl12,styl13,styl13b,nade13,bost14,karp14}, and leads to a failure of   the Johnson-Kendall-Roberts theory of adhesion \cite{john71,styl13c,sale13,xu14,cao14}.
Of particular relevance are our recent experiments embedding droplets in soft solids, where we found that Eshelby's predictions could not describe the response of inclusions below a critical, micron-scale elastocapillary length \cite{styl14b}.
A similar break down was also seen in recent experiments that embedded bubbles in soft, elastic foams \cite{ducl14}.

To apply Eshelby's theory to a broad-class of mechanical phenomena  in soft materials, we need to reformulate it to account for surface tension.
Here, we derive analytic expressions for the deformation of individual inclusions, the deformation and stress fields around the inclusions, and the elastic moduli of soft composites.
Our approach builds upon previous theoretical works that have: focused on strain-dependent surface stresses \cite{duan05,duan07a,duan07b,bris10,bris10a} (which are relevant to nanoinclusions in stiffer materials, but not for softer materials such as gels \cite{hui13}), only considered isotropic loadings \cite{shar04}, used incorrect boundary conditions \cite{yang04} (cf \cite{duan05b}), or only considered incompressible solids and employed a dipole approximation to calculate composite properties \cite{pali90}.

\section{Stretching individual inclusions \label{sec:individ}}

\begin{figure}
\centering
\includegraphics[width=6cm]{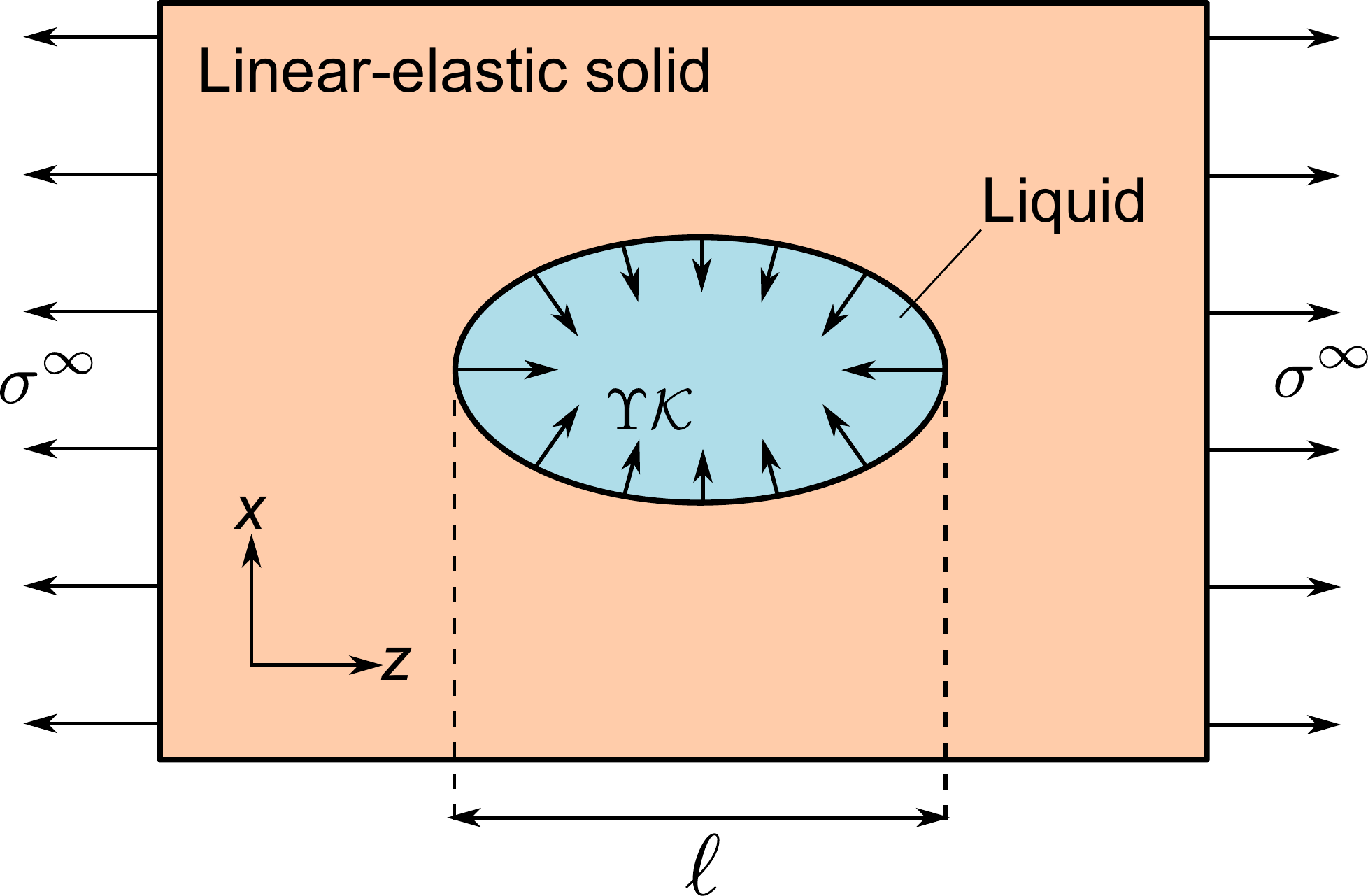}
  \caption{Schematic diagram for uniaxial stretching of a single, incompressible droplet embedded in a linear-elastic solid. $\ell$ is the length of the deformed droplet in the stretch direction.}
  \label{fig:schematic}
\end{figure}

We begin by considering how surface tension affects Eshelby's solution for the deformation of individual inclusions embedded in elastic solids subjected to far-field stresses \cite{eshe57}.
We consider an isolated, incompressible, spherical droplet of radius $R$ embedded in a linear-elastic solid that is deformed by a constant uniaxial far-field stress, as shown in Figure \ref{fig:schematic}.
The displacement field in the solid satisfies
\begin{equation}
\label{eqn:elasticity}
(1-2\nu)\nabla^2\mathbf{u}+ \nabla(\nabla\cdot\mathbf{u})=0,
\end{equation}
where $\nu$ is Poisson's ratio of the solid.

For far-field boundary conditions, the stress in the solid $\mathbf{\sigma}$ is given by the applied uniaxial stress $\sigma_{zz}=\sigma^\infty$, $\sigma_{xx}=\sigma_{yy}=0$ in cartesian coordinates.
Stress and strain are related by
\begin{equation}
\label{eqn:constit_eqn}
\epsilon_{ij}=\frac{1}{E}\left[(1+\nu)\sigma_{ij}-\nu \delta_{ij} \sigma_{kk} \right],
\end{equation}
where $\delta_{ij}$ is the Kronecker delta, and $E$ is Young's modulus.
Thus, the far-field boundary conditions can also be written $\epsilon_{zz}=\epsilon_{zz}^\infty=\sigma_{zz}^\infty/E$, $\epsilon_{xx}=\epsilon_{yy}=-\nu\epsilon_{zz}^\infty$.
At the surface of the droplet the elastic stress satisfies a generalised Young-Laplace equation, which states that the difference in normal stress across an interface depends on its surface stress, $\Upsilon$, and curvature ${\cal K}$ (equal to twice the mean curvature, or the sum of the principal curvatures) via
\begin{equation}
\label{eqn:stressbc}
\sigma\cdot\mathbf{n}=-p\mathbf{n}+\Upsilon {\cal K}\mathbf{n}
\end{equation}
(e.g. \cite{mora11,styl12}).
Here $\mathbf{n}$ is the normal to the deformed droplet surface, $\sigma\cdot\mathbf{n}$ is the normal stress on the solid side, and $p$ is the pressure in the droplet.
The assumption that the surface stress is simply an isotropic, strain-independent, surface tension is appropriate for many soft materials including gels and elastomers \cite{hui13}.
Expressions for $\mathbf{n}$ and ${\cal K}$ in terms of surface displacements are given in the Appendix -- these are different from the expressions used in \cite{yang04} which ignored inclusion deformation and assumed that ${\cal K}=2\Upsilon/R$ \cite{duan05}.

We exploit symmetry and solve the problem in spherical polar co-ordinates by adopting as an ansatz the solution
\begin{multline*}
u_r= {\cal F}r + \frac{{\cal G}}{r^2}+{\cal P}_2(\cos\theta)\times \\
\left[12 \nu {\cal A}r^3+2{\cal B}r+2\frac{(5-4\nu){\cal C}}{r^2}-3\frac{{\cal D}}{r^4} \right],
\end{multline*}
and
\begin{multline}
\label{eqn:u}
u_\theta=\frac{d {\cal P}_2(\cos\theta)}{d\theta}\times \\
\left[(7-4\nu) {\cal A}r^3+{\cal B}r+2\frac{(1-2\nu){\cal C}}{r^2}+\frac{{\cal D}}{r^4} \right].
\end{multline}
as described by \cite{duan05}.  The surface displacements in the radial and $\theta$ directions ($\theta$ is the polar angle from the $z$-axis) are
$u_r$ and $u_\theta$ respectively, ${\cal P}_2$ is the Legendre polynomial of order 2, and ${\cal A}$ through ${\cal G}$ will be determined from the boundary conditions.

Applying the far-field strain condition, we find that ${\cal A}=0$, ${\cal F}=(1-2\nu)\epsilon_{zz}^\infty/3$ and ${\cal B}=(1+\nu)\epsilon_{zz}^\infty/3$. 
Droplet incompressibility requires that $\int_{\cal S} \mathbf{u}\cdot\mathbf{n}\, dS=\int_{0}^{2\pi}\int_0^\pi R^2 u_r\sin\theta \,d\theta d\phi=0$, where ${\cal S}$ is the boundary of the stretched droplet, and the area integral is evaluated using results from differential geometry summarized in the Appendix.
This gives ${\cal G}=-(1-2\nu)R^3\epsilon_{zz}^\infty/3$.  Finally, by applying the boundary condition (\ref{eqn:stressbc}) using equation (\ref{eqn:constit_eqn}) to covert stresses to strains and displacements we obtain 
\begin{equation}
{\cal C}=\frac{5R^3(1+\nu)[\frac{R}{L}-(1+\nu)]}{6\left[\frac{R}{L}(7-5\nu)+(17 -2\nu -19\nu^2)\right]}\epsilon_{zz}^\infty
\end{equation}
and
\begin{equation}
{\cal D}=\frac{R^5(1+\nu)[\frac{R}{L}-(-1+\nu+2\nu^2)]}{\frac{R}{L}(7-5\nu)+(17 -2\nu -19\nu^2)}\epsilon_{zz}^\infty.
\end{equation}
Here $L\equiv\Upsilon/E$ is the elastocapillary length, a material property of the solid/liquid interface. For perturbations of wavelength much smaller than $L$, $\lambda \ll L$, surface deformations are primarily opposed by surface tension, whereas for $\lambda \gg L$, bulk elasticity suppresses deformation of the surface  (e.g. \cite{wang06,mora10,styl13b,styl14}).
\begin{figure*}
\centering
\includegraphics[width=15cm]{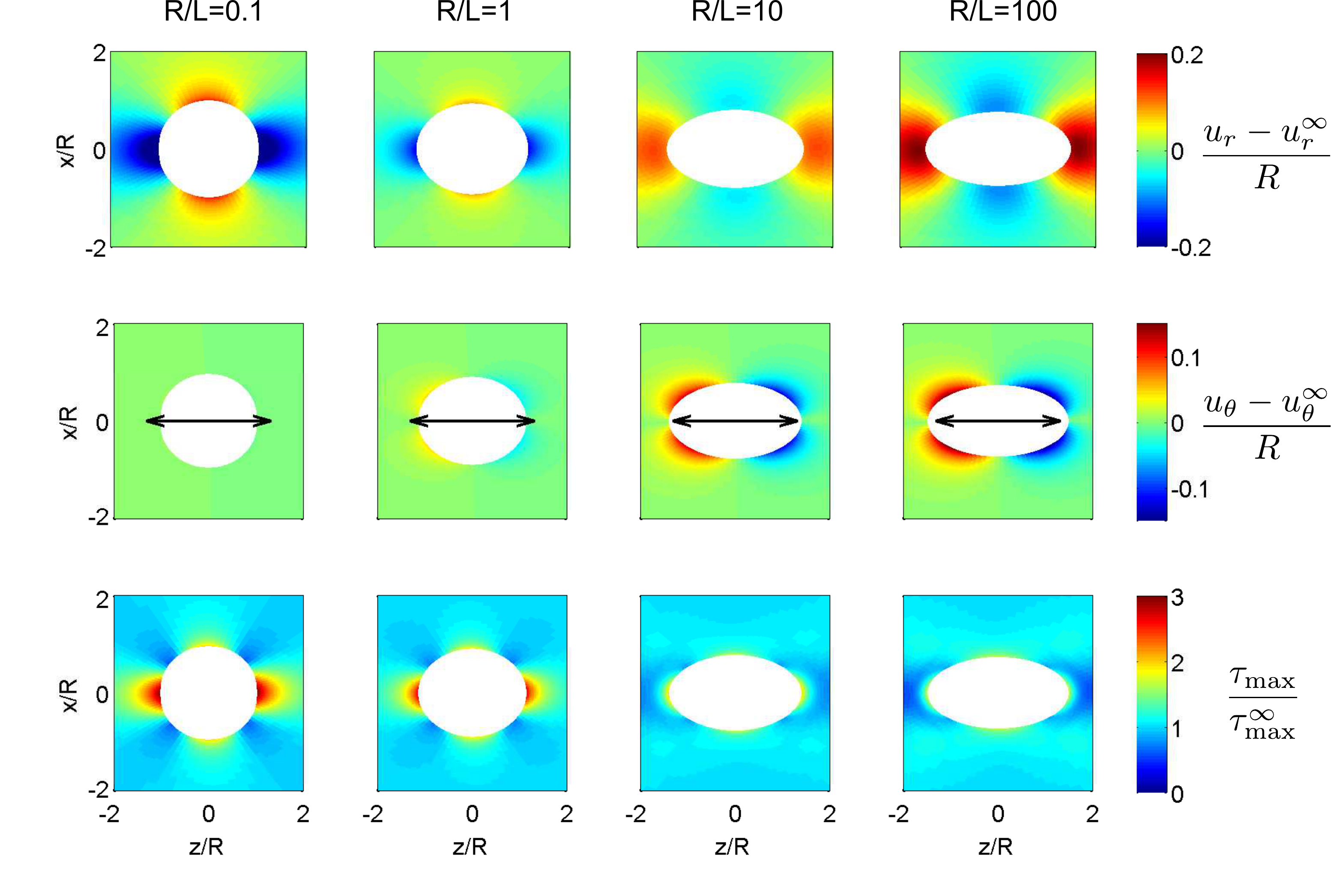}
  \caption{Examples of droplets embedded in an incompressible solid under uniaxial strain with $\epsilon_{zz}^\infty=0.3$. Top: excess radial displacements $(u_r-u_r^\infty)/R$ caused by the presence of the inclusion. The elastic dipole around the inclusion changes sign as $R/L$ increases. Middle: excess tangential displacements $(u_\theta-u_\theta^\infty)/R$. $\theta$ is the polar angle from the $z$-axis. Bottom: shear-stress concentration factor $\tau_\mathrm{max}/\tau_\mathrm{max}^\infty$. When surface tension dominates, $\tau_\mathrm{max}$ is significantly increased at the inclusion tip. The black arrows denote the stretch of the host material.}
  \label{fig:heat_maps}
\end{figure*}
With the expressions for ${\cal A-G}$, Equation (\ref{eqn:u}) gives us the exact displacement solutions.
These also allow calculation of stresses in the solid: we convert displacements to strains (e.g. \cite{land86}) and then use Equation (\ref{eqn:constit_eqn}) to find the non-zero stress components $\sigma_{rr}$, $\sigma_{r\theta}$, $\sigma_{\theta \theta}$ and $\sigma_{\phi \phi}$.

While these results are for uniaxial stress, they are readily extended to provide the solution for general far-field stresses.
In the appropriate coordinate frame, the far-field stress matrix is diagonalisable so the only non-zero far-field stresses are $\sigma_{1}$, $\sigma_{2}$ and $\sigma_{3}$. Then, from linearity of the governing equations, we can calculate the resulting displacements by simply summing the solutions for uniaxial far-field stresses $\sigma_{1}$, $\sigma_{2}$ and $\sigma_{3}$.

\subsection{Inclusion shape}

While Eshelby's results are scale-free, surface tension makes the response of a liquid inclusion strongly size-dependent
For large droplets, $R \gg L$, the fluid droplet deforms more than the surrounding solid.  
In this limit,  the droplet shape only depends on $\sigma^\infty/E$ and $\nu$, in agreement with Eshelby's theory \cite{eshe57}.
However, as $R$ approaches $L$, high  interfacial curvatures are suppressed by surface tension.  For $R\ll L$,  $u_r(R,\theta)=0$ and the droplets remain spherical.
This is visualized in Figure \ref{fig:heat_maps} for a uniaxially-stressed solid where the far-field stress and strain are $\sigma^\infty=0.3E$ and $\epsilon_{zz}^\infty=0.3$, respectively.
Here, the radial and polar displacements in the top two rows are measured relative to the far-field displacement field: $u_r^\infty={\cal F}r+2P_2(\cos\theta){\cal B}r$ and $u_\theta^\infty=P_2'(\cos\theta){\cal B}r$.

\begin{figure}
\centering
\includegraphics[width=9cm]{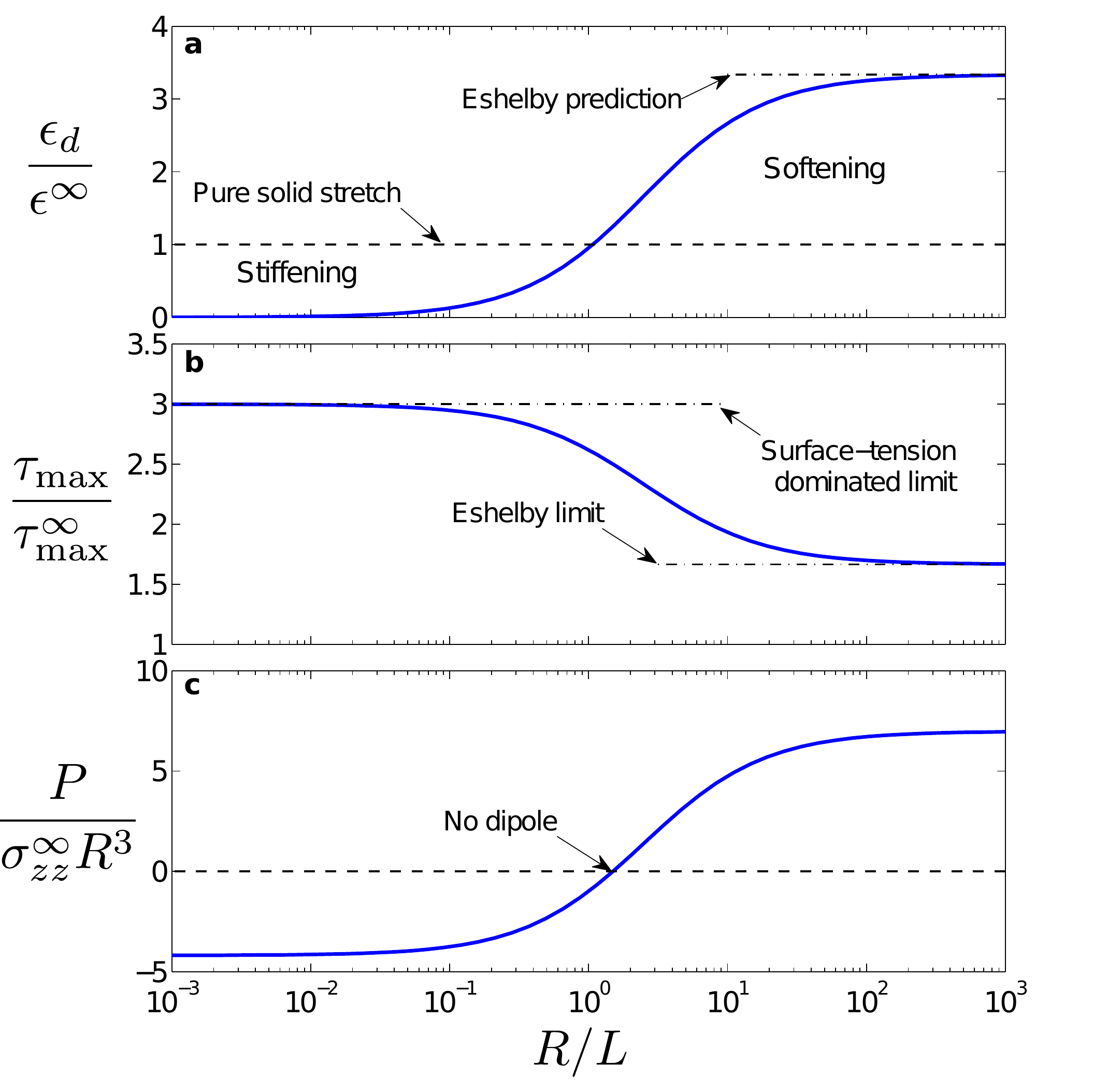}
  \caption{Liquid inclusion characteristics as a function of size $R/L$ for inclusions in an incompressible solid with an applied uniaxial far-field stress as shown in Figure \ref{fig:schematic}. a) Droplet strain, $\epsilon_d=(\ell-2R)/R$ divided by far-field strain $\epsilon^\infty$ only depends on $R/L$. When $ R/L \ll 1$, surface tension dominates and there is no droplet deformation. When $ R/L \gg 1$, surface tension is negligible and the shape prediction is that of classical Eshelby theory -- given by the dash-dotted line. The dashed line shows the material stretch, $(\ell-2R)/R=\epsilon^\infty$.
b) The shear-stress concentration factor at the inclusion tip ($r=R,\theta=0$). This corresponds to the highest shear stress in the solid around the inclusion (see Figure \ref{fig:heat_maps}, bottom row). Dash-dotted lines show  surface-tension dominated and Eshelby limits: $\tau_\mathrm{max}/\tau_\mathrm{max}^\infty=3,\, 5/3$ respectively.
c) The far-field dipole caused around the inclusion. Note that this dipole changes sign at $R=1.5L$, indicating the transition between inclusion stiffening and inclusion softening of the composite.
}
  \label{fig:RE_gamma_figs}
\end{figure}

Changes in droplet shape are captured with an effective droplet strain $\epsilon_d=(\ell-2R)/R$, where $\ell$ is the long-axis of the droplet.
For an incompressible solid, Eq. \ref{eqn:u} gives
\begin{equation}
\epsilon_d=\left(\frac{20\epsilon_{zz}^\infty}{6+15\frac{L}{R}}\right).
\end{equation}
This is plotted in Figure \ref{fig:RE_gamma_figs}(a).
In both extremes of droplet size,  the droplet shape is independent of size.  
In the capillary-dominated regime  ($R\ll L$) the droplet stays spherical ($\epsilon_d=0$).
In the large-droplet limit ($R\gg L$), surface tension does not play a role, and Eshelby's results are recovered ($\epsilon_d=10\epsilon_{xx}^\infty/3$).
There is a smooth cross-over between these limits in the vicinity of $R \sim L$.  
Surface tension makes significant changes to droplet shape for droplet radii up to about $100L$.

Although we only consider the uniaxial stress case above, similar results are obtained for more general triaxial far-field stresses.
For example, for an incompressible solid in plane stress conditions ($\sigma_1,\sigma_2\neq 0$, $\sigma_{3}^\infty=0$, e.g. \cite{styl14b}), the length of the droplet in the $1$-direction is
\begin{equation}
\ell_1=2R\left[1+\frac{5(2\sigma_{1}^\infty-\sigma_{2}^\infty)}{E\left(6+15\frac{L}{R}\right)} \right].
\end{equation}
We recently compared this result to experimental measurements of individual liquid inclusions in soft, stretched solids.
We found fairly good agreement over a wide range of droplet sizes, substrate stiffnesses and applied strains \cite{styl14b}.

\subsection{Stress focussing by inclusions}

The macroscopic strength of composites can be reduced due to stress focusing by inclusions.  
According to the Tresca yield condition, the solid will yield when the shear stress exceeds a critical value $\tau_c$.
Figure \ref{fig:heat_maps} (bottom row) shows the maximum shear stress, $\tau_\mathrm{max}$, for an incompressible solid with a uniaxial far-field stress for various values of $R/L$.
The maximum shear stress is greatest at the tip of the inclusion, and the value there increases significantly as $R$ is reduced below $L$.
In fact at the inclusion tip
\begin{equation}
\label{eqn:tau}
\tau_\mathrm{max}(r=R,\theta=0)=\tau_\mathrm{max}^\infty\frac{5(2+9\frac{L}{R})}{6 +15\frac{L}{R}}.
\end{equation}
This is plotted in Figure \ref{fig:RE_gamma_figs}(b).
There is a significant increase in shear-stress concentration as surface tension becomes more important with
$\tau_\mathrm{max}(R,0)$ increasing from $5\tau_\mathrm{max}^\infty/3$ when $R\gg L$, to $3\tau_\mathrm{max}^\infty$ when $R\ll L$.

These results suggest that surface tension could weaken a soft composite when inclusions fall below a size $\sim 100L$.
This also means that the applied strain at which yielding is expected to occur is no longer independent of the size of the liquid inclusion, as would be predicted from Eshelby's results, but depends on the parameter $R/L$.
These results hint at the potential role of surface tension for fracture mechanics in soft materials where a critical value is the crack-tip stress.
The capillary-induced stress focussing seen here shows how surface tension could potentially significantly alter this value \cite{liu14}. 

\subsection{Dipole signature of inclusions}

At finite concentrations, inclusions interact at a distance through their far-field stresses.  
This can be important for determining mechanical properties of dilute composites (e.g. \cite{pali90,schw13}).  
The far-field solutions are conveniently expressed by a multipole expansion.  

Here,  inclusions appear as force dipoles in the far-field.
From Equations (\ref{eqn:u}), we find the leading order terms in the inclusion-induced displacements ($u_r-u_r^\infty$, $u_\theta-u_\theta^\infty$) are proportional to $1/r^2$.
This corresponds to a force dipole in an elastic body \begin{equation} P_{ij}=P\hat{z}_i\hat{z}_j+P_e\delta_{ij}, \end{equation} with $\mathbf{\hat{z}}$ being the unit vector in the $z$-direction.
The displacement fields due to the dipoles are \cite{luri05}
\begin{equation}
u_r=\frac{(1+\nu)\left[(1-2\nu)\left(P+3P_e\right)+{\cal P}_2(\cos\theta)(5-4\nu)P\right]}{12 \pi E (1-\nu) r^2},
\end{equation}
and
\begin{equation}
u_\theta=\frac{(1+\nu)(1-2\nu)}{12 \pi E r^2(1-\nu)}\frac{d {\cal P}_2(\cos\theta)}{d\theta}P.
\end{equation}
Thus, from comparison with Equation (\ref{eqn:u}),
\begin{equation}
P=24{\cal C}\pi E (1-\nu)/(1+\nu),\end{equation} 
and
\begin{equation} P_e=4\pi E(1-\nu)\frac{{\cal G}-2{\cal C}(1-2\nu)}{(1+\nu)(1-2\nu)}.
\end{equation}
The first dipole, $P$, is a force dipole of two point forces on the $z$-axis which also act along the $z$-axis - i.e. parallel to the applied far-field stress. The second term $P_e$ is an isotropic centre of expansion \cite{luri05}. When $\nu=1/2$, the displacement field due to $P_e$ vanishes, and $P=8\pi {\cal C} E$.

Intriguingly, the dipole strength, $P$, can be positive or negative.
Figure \ref{fig:RE_gamma_figs}(c) shows the normalised dipole strength $P/\sigma^\infty R^3$ of a liquid inclusion in incompressible solid with a uniaxial applied stress.
For large inclusions ($R>1.5 L$), $P>0$ and the dipole is a pair of outward pointing point forces. This increases solid displacements -- consistent with a weak point in the solid. 
For small inclusions ($R<1.5 L$), $P<0$ and so the dipole opposes the applied far-field stress, acting like a stiff point in the solid.
The sign switch is clearly seen in the displacement fields of Figure \ref{fig:heat_maps}.
At $R=1.5L$, the inclusion has no effect on the far-field elasticity field and is effectively invisible (e.g., see \cite{milt06}).

\section{Soft composites}

We have shown that the surface tension of a small liquid droplet in a soft linear elastic solid resists deformation imposed by far-field stretch.  Therefore, we expect that the dispersion of small liquid droplets within a solid can increase its apparent macroscopic stiffness.  
We calculate the effective Young's modulus $E_c$ of a composite containing a dilute quantity of monodisperse droplets by following Eshelby's original approach \cite{eshe57,yang04}.
First, we calculate the excess energy $W$ due to the presence of a single inclusion when a solid is uniaxially stretched.
Then, we consider uniaxial stretching of a dilute composite of noninteracting inclusions.
If the applied stress is $\sigma_{zz}=\sigma^\infty$, the strain energy density of the composite is
\begin{equation}
\label{eqn:sed}
{\cal E}=({\sigma^\infty})^2/(2E)+W\Phi/(4\pi R^3/3) = ({\sigma^\infty})^2/(2E_c),
\end{equation}
where $\Phi$ is the volume fraction of inclusions.
The second equality comes from the relationship between the strain energy density and the effective modulus of the material, allowing calculation of $E_c$ from $W$.

The excess energy due to the presence of a single elastic inclusion in a uniaxially-stressed solid is 
\begin{multline}
W=\frac{1}{2}\int_{V_i}(\sigma_{ij}\epsilon_{ij}-\sigma^\infty_{ij}\epsilon^\infty_{ij}) dV \\
+ \frac{1}{2}\int_{V_m}(\sigma_{ij}\epsilon_{ij}-\sigma^\infty_{ij}\epsilon^\infty_{ij}) dV+ \Upsilon \Delta S.
\end{multline}
Here we assume that the inclusion is an elastic solid for generality -- the droplet is the limiting case of zero shear modulus. 

The volumes of the elastic matrix outside of the inclusion and the inclusion $V_m$ and $V_i$, respectively, the far-field stresses/strains are $\sigma_{ij}^\infty$ and $\epsilon_{ij}^\infty$ respectively, and the change in surface are of the droplet upon stretching is $\Delta S$. 
Using the divergence theorem, the stress boundary condition (\ref{eqn:stressbc}), and the fact that in the far-field, $\sigma_{ij}^\infty=\sigma_{ij}$,
\begin{eqnarray}
\nonumber W=&\frac{1}{2}&\int_{V_m}(\sigma_{ij}^\infty\epsilon_{ij}-\sigma_{ij}\epsilon_{ij}^\infty)dV\\
\nonumber +&\frac{1}{2}&\int_{{\cal S}^+}(n_i\sigma^\infty_{ij}u_j-n_i\sigma_{ij}u^\infty_j)dS\\
-&\frac{\Upsilon}{2}&\int_{\cal S} {\cal K}u_in_i dS+\Upsilon\Delta S \label{eqn:W}.
\end{eqnarray}
Integration on the matrix side of the droplet surface ${\cal S}$ is denoted by ${\cal S}^{+}$.
From Equation (\ref{eqn:constit_eqn}), the first term is zero, so $W$ depends only upon displacements and stresses at the droplet surface.
Using our earlier results (e.g. Equations \ref{eqn:u}), along with second-order (in the displacement) versions of the expressions for $\mathbf{n}$, ${\cal K}$, $dS$ and $\Delta S$ shown in the Appendix, we obtain $W$ for the case of a uniaxial far-field stress $\sigma^\infty$:
\begin{multline}
\label{eqn:W}
W=2\pi R^3{\sigma^\infty}^2(1-\nu) \\
\times \frac{[\frac{R}{L}(1+13\nu)-(9-2\nu+5\nu^2+16\nu^3)]}{E(1+\nu)[\frac{R}{L}(7-5\nu)+(17-2\nu-19\nu^2)]}.
\end{multline}
Finally, from Equation (\ref{eqn:sed}),
\begin{multline}
\frac{E_c}{E}=\\
\left[1+\frac{3(1-\nu)\left[\frac{R}{L}(1+13\nu) - (9-2\nu+5\nu^2+16\nu^3)\right]}{(1+\nu)\left[\frac{R}{L}(7-5\nu)+(17-2\nu-19\nu^2)\right]}\Phi \right]^{-1}
\end{multline}
For an incompressible solid $\nu = 1/2$ and we have
\begin{equation}
\label{eqn:E_c}
\frac{E_c}{E}=\frac{1+\frac{5}{2}\frac{L}{R}}{\frac{5}{2}\frac{L}{R}(1-\Phi)+(1+\frac{5}{3}\Phi)}.
\end{equation}

\begin{figure}
\centering
\includegraphics[width=9cm]{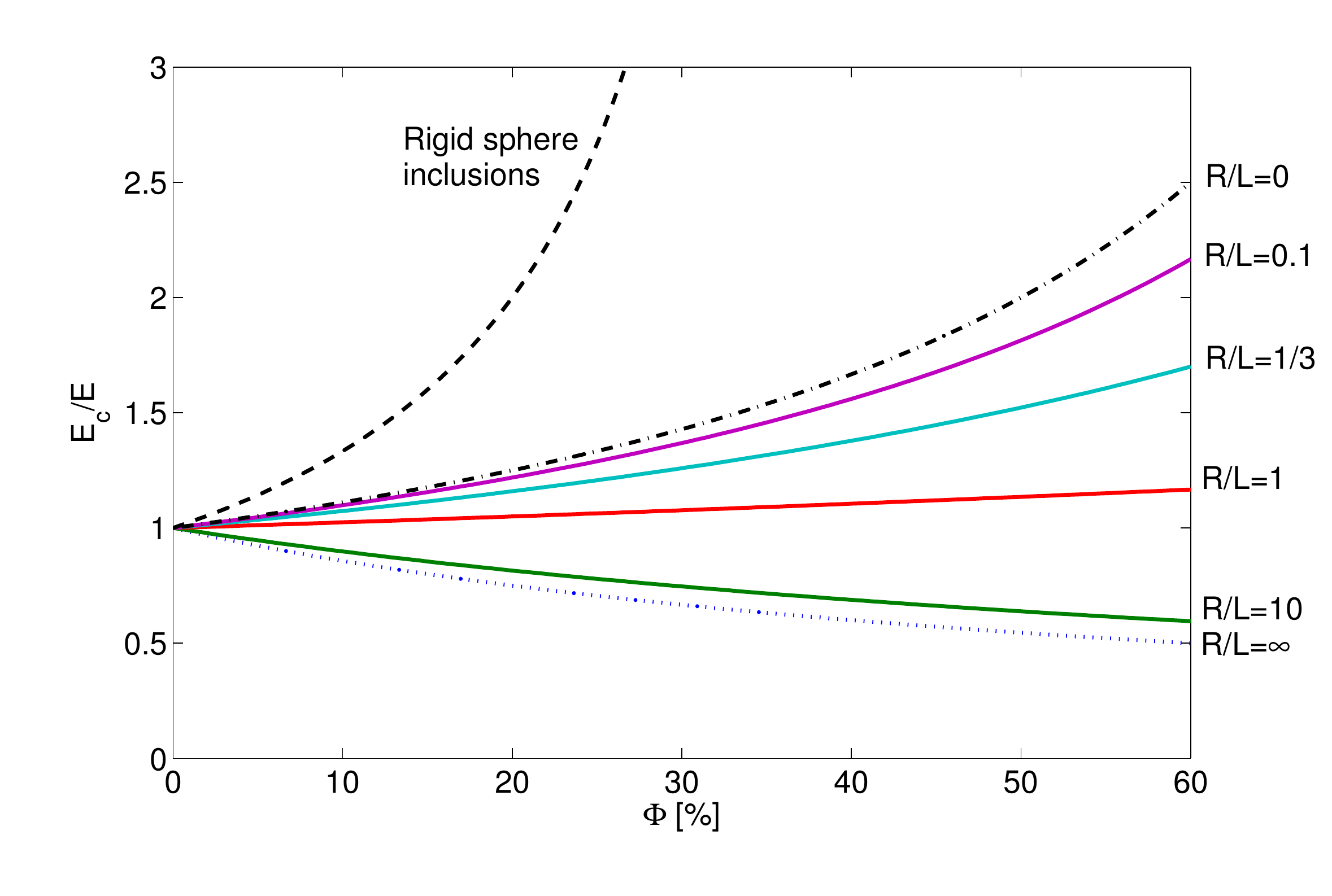}
  \caption{The stiffness of soft composites. Young's modulus of composites of droplets embedded in linear-elastic solids, $E_c$ as a function of liquid content. The dotted curve shows Eshelby's prediction without surface tension. The dash-dotted curve shows the surface-tension dominated limit, $R/L\ll1$. The dashed curve show Eshelby's prediction for rigid spheres embedded in an elastic solid.}
  \label{fig:composites}
\end{figure}

Figure \ref{fig:composites} plots the results of Equation (\ref{eqn:E_c}) and shows the dramatic influence of capillarity on soft composite stiffness.
When surface tension is negligible ($R\gg L$), the composite becomes more compliant as the density of droplets increases -- in exact agreement with Eshelby's prediction of $E_c/E=(1+5\Phi/3)^{-1}$ (dotted curve), and qualitatively agreeing with other classical composite laws (e.g. \cite{hash63,mori73}).
However Eshelby's predictions break down when $R\lesssim 100 L$.
In fact, when $R<1.5L$, increasing the density of droplets causes the solid to stiffen, consistent with the dipole sign-switching seen earlier.
In the surface-tension dominated limit, $R\ll L$, the droplets stay spherical, and we find the maximum achievable composite stiffness $E_c=E/(1-\Phi)$ (dash-dotted curve).
Note that the droplets do not behave like rigid particles in this limit, for which $E_c=E/(1-5\Phi/2)$ \cite{eshe57} (dashed curve). Although the droplets remain spherical due to capillarity, there are non-zero tangential displacements, unlike the case of rigid particles.

These results agree with experiments.
Recently, we made soft composites of glycerol droplets embedded in soft silicone solids.
In quantitative agreement with the theory, we saw stiffening of solids by droplets in compliant solids, and softening in stiffer solids \cite{styl14b}.
In the dilute limit ($\Phi\rightarrow 0$), Equation (\ref{eqn:E_c}) matches with recent theoretical predictions (derived using the dipole approximation for inclusions in incompressible solids) that describe experimental measurements of shear moduli of emulsions containing monodisperse bubbles \cite{ducl14,pali90}.

\section{Conclusions}

We have modified Eshelby's inclusion theory to include surface tension for liquid inclusions in a linear-elastic solid, giving both the microscopic behaviour and the macroscopic effects of inclusions in composites.
We have shown that surface tension stiffens small inclusions, and focusses shear stresses at the inclusion tips. 
Thus composites with small, capillary-dominated inclusions will be stiffer but may be weaker.
This stress-concentration illustrates the potentially strong role of surface tension in the failure of soft-solids, highlighting the relevance of this work to emerging fields like fracture mechanics and plasticity in soft materials (e.g. \cite{kund09,cui11,liu14}).

Inclusions with surface tension can be viewed, at leading order, as elastic dipoles in a solid.
The sign of the dipole captures the stiffening behaviour due to capillarity.
Treating inclusions as dipoles also offers a simplified picture of inclusions that give the interactions between features in elastic bodies, and can streamline calculations of bulk composite properties via standard theories.
The analytic theory presented for bulk composite stiffness, which incorporates the entire elastic field around inclusions, validates the dipole approach by recovering previous results for incompressible materials in the limit of dilute composites \cite{ducl14,pali90}.

Our work is applicable to a wide variety of soft material problems.
Most obviously it can be directly applied to composites comprising soft materials such as gels and elastomers.
As a specific example, we have shown how surface tension effects allow elastic cloaking, with inclusions of size $R=1.5L$ being mechanically invisible.
Our work also has interesting uses in mechanobiology, as biological tissue is predominantly soft.
For example, a recent study embedded droplets in biological tissue and observed their deformations to extract local anisotropic stresses \cite{camp14}.
The coupling between microscopic  and macroscopic stress also plays an important role in the tensional homeostasis of soft tissues \cite{brow98,zeme06}. 
Although we have only considered liquid inclusions here our analysis can be repeated for more general soft composites with elastic inclusions in place of liquid droplets.
In that case, we expect that similar capillary effects to those presented here will be seen whenever $R\lesssim 100 \Upsilon/E_i, 100\Upsilon/E_m$ with $E_i/E_m$ being the inclusion/matrix stiffnesses respectively.

\section{Acknowledgements}

We thank Peter Howell and Alain Goriely for helpful conversations. We are grateful for funding from the National Science Foundation (CBET-1236086) to ERD, the Yale University Bateman Interdepartmental Postdoctoral Fellowship to RWS and the John Simon Guggenheim Foundation, the Swedish Research Council, and a Royal Society Wolfson Research Merit Award to JSW.

\section{Appendix}

To calculate the effect of surface tension on the shape of a droplet embedded in a soft solid, we need expressions for the normal to the droplet surface, its curvature, and surface area in terms of the surface displacements.
We consider an initially spherical droplet with the position of its surface given by $\mathbf{x}=(R,0,0)$, and apply a uniaxial stretch so that $\mathbf{x} \rightarrow \mathbf{x}'=(R+u_r,u_\theta,0)$.
From axisymmetry, the $u_r,u_\theta$ are independent of the angle $\phi$.

We calculate the normal to the droplet surface, $\mathbf{n}$, by taking the cross-product of the surface tangent vectors, $\partial \mathbf{x}'/\partial \theta$ and $\partial \mathbf{x}'/\partial \phi$ \cite{abbe06},
\begin{equation}
\mathbf{n}=\frac{\frac{\partial \mathbf{x}'}{\partial\theta}\wedge\frac{\partial \mathbf{x}'}{\partial\phi}}{\left|\frac{\partial \mathbf{x}'}{\partial\theta}\wedge\frac{\partial \mathbf{x}'}{\partial\phi}\right|},
\end{equation}
with
\begin{eqnarray}
\partial \mathbf{x}'/\partial \theta&=&\left(\frac{\partial u_r}{\partial \theta}-u_\theta, R+u_r+\frac{\partial u_\theta}{\partial \theta},0\right),
\end{eqnarray}
and
\begin{eqnarray}
\partial \mathbf{x}'/\partial \phi&=&\left(0,0,(R+u_r)\sin\theta +u_\theta\cos\theta \right).
\end{eqnarray}
At leading order in $\mathbf{u}$ we find
\begin{equation}
\label{eqn:norm}
\mathbf{n}=\left(1,\frac{u_\theta}{R}-\frac{1}{R}\frac{\partial u_r}{\partial \theta},0\right).
\end{equation}

The droplet surface curvature, ${\cal K}$, can be calculated from differential geometry using the first and second fundamental forms \cite{abbe06}:
\begin{equation}
{\cal K}=\frac{e_f G_f-2f_fF_f+g_fE_f}{E_fG_f-F_f^2}
\end{equation}
where
\begin{equation}
E_f=\frac{\partial \mathbf{x}'}{\partial \theta}\cdot\frac{\partial \mathbf{x}'}{\partial \theta},\quad F_f=\frac{\partial \mathbf{x}'}{\partial \theta}\cdot\frac{\partial \mathbf{x}'}{\partial \phi},\quad G_f=\frac{\partial \mathbf{x}'}{\partial \phi}\cdot\frac{\partial \mathbf{x}'}{\partial \phi},
\end{equation}
and
\begin{equation}
e_f=\mathbf{n}\cdot\frac{\partial^2 \mathbf{x}'}{\partial \theta^2},\quad f_f=\mathbf{n}\cdot\frac{\partial^2 \mathbf{x}'}{\partial \theta\partial \phi},\quad g_f=\mathbf{n}\cdot\frac{\partial^2 \mathbf{x}'}{\partial \phi^2}.
\end{equation}
Thus, at leading order in $\mathbf{u}$,
\begin{equation}
\label{eqn:curv}
{\cal K} = \frac{2}{R} - \frac{1}{R^2} \left( 2 u_r+\cot\theta \frac{\partial u_r}{\partial \theta}+\frac{\partial^2 u_r}{\partial \theta^2}\right).
\end{equation}

Using the results above, we also obtain the area element $dS=\sqrt{E_fG_f-F_f^2}d\theta d\phi$ \cite{abbe06}.
At leading order in $\mathbf{u}$,
\begin{multline}
\label{eqn:area}
dS=\left[R^2\sin\theta\right. \\
+\left. R\left(u_\theta\cos\theta + 2 u_r\sin\theta+\frac{\partial u_\theta}{\partial \theta}\sin\theta\right)\right]d\theta d\Phi,
\end{multline}
and after integration we obtain the droplet surface area
\begin{multline}
\label{eqn:tot_area}
S=4\pi R^2 \\
+\int_0^{2\pi} \int_0^\pi \left[R\left(u_\theta\cos\theta + 2 u_r\sin\theta+\frac{\partial u_\theta}{\partial \theta}\sin\theta\right)\right]d\theta d\Phi\\
=4\pi R^2 +\int_0^{2\pi} \int_0^\pi 2 u_r\sin\theta\, d\theta d\Phi.
\end{multline}


\begin{thebibliography}{55}%
\makeatletter
\providecommand \@ifxundefined [1]{%
 \@ifx{#1\undefined}
}%
\providecommand \@ifnum [1]{%
 \ifnum #1\expandafter \@firstoftwo
 \else \expandafter \@secondoftwo
 \fi
}%
\providecommand \@ifx [1]{%
 \ifx #1\expandafter \@firstoftwo
 \else \expandafter \@secondoftwo
 \fi
}%
\providecommand \natexlab [1]{#1}%
\providecommand \enquote  [1]{``#1''}%
\providecommand \bibnamefont  [1]{#1}%
\providecommand \bibfnamefont [1]{#1}%
\providecommand \citenamefont [1]{#1}%
\providecommand \href@noop [0]{\@secondoftwo}%
\providecommand \href [0]{\begingroup \@sanitize@url \@href}%
\providecommand \@href[1]{\@@startlink{#1}\@@href}%
\providecommand \@@href[1]{\endgroup#1\@@endlink}%
\providecommand \@sanitize@url [0]{\catcode `\\12\catcode `\$12\catcode
  `\&12\catcode `\#12\catcode `\^12\catcode `\_12\catcode `\%12\relax}%
\providecommand \@@startlink[1]{}%
\providecommand \@@endlink[0]{}%
\providecommand \url  [0]{\begingroup\@sanitize@url \@url }%
\providecommand \@url [1]{\endgroup\@href {#1}{\urlprefix }}%
\providecommand \urlprefix  [0]{URL }%
\providecommand \Eprint [0]{\href }%
\providecommand \doibase [0]{http://dx.doi.org/}%
\providecommand \selectlanguage [0]{\@gobble}%
\providecommand \bibinfo  [0]{\@secondoftwo}%
\providecommand \bibfield  [0]{\@secondoftwo}%
\providecommand \translation [1]{[#1]}%
\providecommand \BibitemOpen [0]{}%
\providecommand \bibitemStop [0]{}%
\providecommand \bibitemNoStop [0]{.\EOS\space}%
\providecommand \EOS [0]{\spacefactor3000\relax}%
\providecommand \BibitemShut  [1]{\csname bibitem#1\endcsname}%
\let\auto@bib@innerbib\@empty
\bibitem [{\citenamefont {Eshelby}(1957)}]{eshe57}%
  \BibitemOpen
  \bibfield  {author} {\bibinfo {author} {\bibfnamefont {J.~D.}\ \bibnamefont
  {Eshelby}},\ }\href@noop {} {\bibfield  {journal} {\bibinfo  {journal} {Proc.
  Roy. Soc. Lond. A}\ }\textbf {\bibinfo {volume} {241}},\ \bibinfo {pages}
  {376} (\bibinfo {year} {1957})}\BibitemShut {NoStop}%
\bibitem [{\citenamefont {Hashin}\ and\ \citenamefont
  {Shtrikman}(1963)}]{hash63}%
  \BibitemOpen
  \bibfield  {author} {\bibinfo {author} {\bibfnamefont {Z.}~\bibnamefont
  {Hashin}}\ and\ \bibinfo {author} {\bibfnamefont {S.}~\bibnamefont
  {Shtrikman}},\ }\href@noop {} {\bibfield  {journal} {\bibinfo  {journal} {J.
  Mech. Phys. Solids}\ }\textbf {\bibinfo {volume} {11}},\ \bibinfo {pages}
  {127} (\bibinfo {year} {1963})}\BibitemShut {NoStop}%
\bibitem [{\citenamefont {Mori}\ and\ \citenamefont {Tanaka}(1973)}]{mori73}%
  \BibitemOpen
  \bibfield  {author} {\bibinfo {author} {\bibfnamefont {T.}~\bibnamefont
  {Mori}}\ and\ \bibinfo {author} {\bibfnamefont {K.}~\bibnamefont {Tanaka}},\
  }\href@noop {} {\bibfield  {journal} {\bibinfo  {journal} {Acta Metall.}\
  }\textbf {\bibinfo {volume} {21}},\ \bibinfo {pages} {571 } (\bibinfo {year}
  {1973})}\BibitemShut {NoStop}%
\bibitem [{\citenamefont {Hill}(1963)}]{hill63}%
  \BibitemOpen
  \bibfield  {author} {\bibinfo {author} {\bibfnamefont {R.}~\bibnamefont
  {Hill}},\ }\href@noop {} {\bibfield  {journal} {\bibinfo  {journal} {J. Mech.
  Phys. Solids}\ }\textbf {\bibinfo {volume} {11}},\ \bibinfo {pages} {357}
  (\bibinfo {year} {1963})}\BibitemShut {NoStop}%
\bibitem [{\citenamefont {Budiansky}(1965)}]{budi65}%
  \BibitemOpen
  \bibfield  {author} {\bibinfo {author} {\bibfnamefont {B.}~\bibnamefont
  {Budiansky}},\ }\href@noop {} {\bibfield  {journal} {\bibinfo  {journal} {J.
  Mech. Phys. Solids}\ }\textbf {\bibinfo {volume} {13}},\ \bibinfo {pages}
  {223} (\bibinfo {year} {1965})}\BibitemShut {NoStop}%
\bibitem [{\citenamefont {Rice}(1968)}]{rice68}%
  \BibitemOpen
  \bibfield  {author} {\bibinfo {author} {\bibfnamefont {J.~R.}\ \bibnamefont
  {Rice}},\ }\href@noop {} {\bibfield  {journal} {\bibinfo  {journal} {J. Appl.
  Mech.}\ }\textbf {\bibinfo {volume} {35}},\ \bibinfo {pages} {379} (\bibinfo
  {year} {1968})}\BibitemShut {NoStop}%
\bibitem [{\citenamefont {Budiansky}\ and\ \citenamefont
  {O'Connell}(1976)}]{budi76}%
  \BibitemOpen
  \bibfield  {author} {\bibinfo {author} {\bibfnamefont {B.}~\bibnamefont
  {Budiansky}}\ and\ \bibinfo {author} {\bibfnamefont {R.~J.}\ \bibnamefont
  {O'Connell}},\ }\href@noop {} {\bibfield  {journal} {\bibinfo  {journal}
  {Int. J. Solids Struct.}\ }\textbf {\bibinfo {volume} {12}},\ \bibinfo
  {pages} {81} (\bibinfo {year} {1976})}\BibitemShut {NoStop}%
\bibitem [{\citenamefont {Mura}(1987)}]{mura87}%
  \BibitemOpen
  \bibfield  {author} {\bibinfo {author} {\bibfnamefont {T.}~\bibnamefont
  {Mura}},\ }\href@noop {} {\emph {\bibinfo {title} {Micromechanics of Defects
  in Solids}}},\ Vol.~\bibinfo {volume} {3}\ (\bibinfo  {publisher}
  {Springer},\ \bibinfo {year} {1987})\BibitemShut {NoStop}%
\bibitem [{\citenamefont {Hutchinson}(1970)}]{hutc70}%
  \BibitemOpen
  \bibfield  {author} {\bibinfo {author} {\bibfnamefont {J.}~\bibnamefont
  {Hutchinson}},\ }\href@noop {} {\bibfield  {journal} {\bibinfo  {journal}
  {Proc. Roy. Soc. London A}\ }\textbf {\bibinfo {volume} {319}},\ \bibinfo
  {pages} {247} (\bibinfo {year} {1970})}\BibitemShut {NoStop}%
\bibitem [{\citenamefont {Berveiller}\ and\ \citenamefont
  {Zaoui}(1978)}]{berv78}%
  \BibitemOpen
  \bibfield  {author} {\bibinfo {author} {\bibfnamefont {M.}~\bibnamefont
  {Berveiller}}\ and\ \bibinfo {author} {\bibfnamefont {A.}~\bibnamefont
  {Zaoui}},\ }\href@noop {} {\bibfield  {journal} {\bibinfo  {journal} {J.
  Mech. Phys. Solids}\ }\textbf {\bibinfo {volume} {26}},\ \bibinfo {pages}
  {325} (\bibinfo {year} {1978})}\BibitemShut {NoStop}%
\bibitem [{\citenamefont {Kanamori}\ and\ \citenamefont
  {Anderson}(1975)}]{kana75}%
  \BibitemOpen
  \bibfield  {author} {\bibinfo {author} {\bibfnamefont {H.}~\bibnamefont
  {Kanamori}}\ and\ \bibinfo {author} {\bibfnamefont {D.~L.}\ \bibnamefont
  {Anderson}},\ }\href@noop {} {\bibfield  {journal} {\bibinfo  {journal}
  {Bull. Seismol. Soc. Am.}\ }\textbf {\bibinfo {volume} {65}},\ \bibinfo
  {pages} {1073} (\bibinfo {year} {1975})}\BibitemShut {NoStop}%
\bibitem [{\citenamefont {Sharma}\ and\ \citenamefont {Ganti}(2004)}]{shar04}%
  \BibitemOpen
  \bibfield  {author} {\bibinfo {author} {\bibfnamefont {P.}~\bibnamefont
  {Sharma}}\ and\ \bibinfo {author} {\bibfnamefont {S.}~\bibnamefont {Ganti}},\
  }\href@noop {} {\bibfield  {journal} {\bibinfo  {journal} {J. Appl. Mech.}\
  }\textbf {\bibinfo {volume} {71}},\ \bibinfo {pages} {663} (\bibinfo {year}
  {2004})}\BibitemShut {NoStop}%
\bibitem [{\citenamefont {Yang}(2004)}]{yang04}%
  \BibitemOpen
  \bibfield  {author} {\bibinfo {author} {\bibfnamefont {F.}~\bibnamefont
  {Yang}},\ }\href@noop {} {\bibfield  {journal} {\bibinfo  {journal} {J. Appl.
  Phys.}\ }\textbf {\bibinfo {volume} {95}},\ \bibinfo {pages} {3516} (\bibinfo
  {year} {2004})}\BibitemShut {NoStop}%
\bibitem [{\citenamefont {Duan}\ \emph
  {et~al.}(2005{\natexlab{a}})\citenamefont {Duan}, \citenamefont {Wang},
  \citenamefont {Huang},\ and\ \citenamefont {Karihaloo}}]{duan05}%
  \BibitemOpen
  \bibfield  {author} {\bibinfo {author} {\bibfnamefont {H.~L.}\ \bibnamefont
  {Duan}}, \bibinfo {author} {\bibfnamefont {J.}~\bibnamefont {Wang}}, \bibinfo
  {author} {\bibfnamefont {Z.~P.}\ \bibnamefont {Huang}}, \ and\ \bibinfo
  {author} {\bibfnamefont {B.~L.}\ \bibnamefont {Karihaloo}},\ }\href@noop {}
  {\bibfield  {journal} {\bibinfo  {journal} {Proc. Roy. Soc. A}\ }\textbf
  {\bibinfo {volume} {461}},\ \bibinfo {pages} {3335} (\bibinfo {year}
  {2005}{\natexlab{a}})}\BibitemShut {NoStop}%
\bibitem [{\citenamefont {Hui}\ \emph {et~al.}(2002)\citenamefont {Hui},
  \citenamefont {Jagota}, \citenamefont {Lin},\ and\ \citenamefont
  {Kramer}}]{hui02}%
  \BibitemOpen
  \bibfield  {author} {\bibinfo {author} {\bibfnamefont {C.~Y.}\ \bibnamefont
  {Hui}}, \bibinfo {author} {\bibfnamefont {A.}~\bibnamefont {Jagota}},
  \bibinfo {author} {\bibfnamefont {Y.~Y.}\ \bibnamefont {Lin}}, \ and\
  \bibinfo {author} {\bibfnamefont {E.~J.}\ \bibnamefont {Kramer}},\
  }\href@noop {} {\bibfield  {journal} {\bibinfo  {journal} {Langmuir}\
  }\textbf {\bibinfo {volume} {18}},\ \bibinfo {pages} {1394} (\bibinfo {year}
  {2002})}\BibitemShut {NoStop}%
\bibitem [{\citenamefont {Jagota}\ \emph {et~al.}(2012)\citenamefont {Jagota},
  \citenamefont {Paretkar},\ and\ \citenamefont {Ghatak}}]{jago12}%
  \BibitemOpen
  \bibfield  {author} {\bibinfo {author} {\bibfnamefont {A.}~\bibnamefont
  {Jagota}}, \bibinfo {author} {\bibfnamefont {D.}~\bibnamefont {Paretkar}}, \
  and\ \bibinfo {author} {\bibfnamefont {A.}~\bibnamefont {Ghatak}},\
  }\href@noop {} {\bibfield  {journal} {\bibinfo  {journal} {Phys. Rev. E}\
  }\textbf {\bibinfo {volume} {85}},\ \bibinfo {pages} {051602} (\bibinfo
  {year} {2012})}\BibitemShut {NoStop}%
\bibitem [{\citenamefont {Mora}\ \emph {et~al.}(2013)\citenamefont {Mora},
  \citenamefont {Maurini}, \citenamefont {Phou}, \citenamefont {Fromental},
  \citenamefont {Audoly},\ and\ \citenamefont {Pomeau}}]{mora13}%
  \BibitemOpen
  \bibfield  {author} {\bibinfo {author} {\bibfnamefont {S.}~\bibnamefont
  {Mora}}, \bibinfo {author} {\bibfnamefont {C.}~\bibnamefont {Maurini}},
  \bibinfo {author} {\bibfnamefont {T.}~\bibnamefont {Phou}}, \bibinfo {author}
  {\bibfnamefont {J.-M.}\ \bibnamefont {Fromental}}, \bibinfo {author}
  {\bibfnamefont {B.}~\bibnamefont {Audoly}}, \ and\ \bibinfo {author}
  {\bibfnamefont {Y.}~\bibnamefont {Pomeau}},\ }\href@noop {} {\bibfield
  {journal} {\bibinfo  {journal} {Phys. Rev. Lett.}\ }\textbf {\bibinfo
  {volume} {111}},\ \bibinfo {pages} {114301} (\bibinfo {year}
  {2013})}\BibitemShut {NoStop}%
\bibitem [{\citenamefont {Paretkar}\ \emph {et~al.}(2014)\citenamefont
  {Paretkar}, \citenamefont {Xu}, \citenamefont {Hui},\ and\ \citenamefont
  {Jagota}}]{pare14}%
  \BibitemOpen
  \bibfield  {author} {\bibinfo {author} {\bibfnamefont {D.}~\bibnamefont
  {Paretkar}}, \bibinfo {author} {\bibfnamefont {X.}~\bibnamefont {Xu}},
  \bibinfo {author} {\bibfnamefont {C.-Y.}\ \bibnamefont {Hui}}, \ and\
  \bibinfo {author} {\bibfnamefont {A.}~\bibnamefont {Jagota}},\ }\href@noop {}
  {\bibfield  {journal} {\bibinfo  {journal} {Soft Matter}\ }\textbf {\bibinfo
  {volume} {10}},\ \bibinfo {pages} {4084} (\bibinfo {year}
  {2014})}\BibitemShut {NoStop}%
\bibitem [{\citenamefont {Mora}\ \emph {et~al.}(2010)\citenamefont {Mora},
  \citenamefont {Phou}, \citenamefont {Fromental}, \citenamefont {Pismen},\
  and\ \citenamefont {Pomeau}}]{mora10}%
  \BibitemOpen
  \bibfield  {author} {\bibinfo {author} {\bibfnamefont {S.}~\bibnamefont
  {Mora}}, \bibinfo {author} {\bibfnamefont {T.}~\bibnamefont {Phou}}, \bibinfo
  {author} {\bibfnamefont {J.-M.}\ \bibnamefont {Fromental}}, \bibinfo {author}
  {\bibfnamefont {L.~M.}\ \bibnamefont {Pismen}}, \ and\ \bibinfo {author}
  {\bibfnamefont {Y.}~\bibnamefont {Pomeau}},\ }\href@noop {} {\bibfield
  {journal} {\bibinfo  {journal} {Phys. Rev. Lett.}\ }\textbf {\bibinfo
  {volume} {105}},\ \bibinfo {pages} {214301} (\bibinfo {year}
  {2010})}\BibitemShut {NoStop}%
\bibitem [{\citenamefont {Mora}\ \emph {et~al.}(2011)\citenamefont {Mora},
  \citenamefont {Abkarian}, \citenamefont {Tabuteau},\ and\ \citenamefont
  {Pomeau}}]{mora11}%
  \BibitemOpen
  \bibfield  {author} {\bibinfo {author} {\bibfnamefont {S.}~\bibnamefont
  {Mora}}, \bibinfo {author} {\bibfnamefont {M.}~\bibnamefont {Abkarian}},
  \bibinfo {author} {\bibfnamefont {H.}~\bibnamefont {Tabuteau}}, \ and\
  \bibinfo {author} {\bibfnamefont {Y.}~\bibnamefont {Pomeau}},\ }\href@noop {}
  {\bibfield  {journal} {\bibinfo  {journal} {Soft Matter}\ }\textbf {\bibinfo
  {volume} {7}},\ \bibinfo {pages} {10612} (\bibinfo {year}
  {2011})}\BibitemShut {NoStop}%
\bibitem [{\citenamefont {Chakrabarti}\ and\ \citenamefont
  {Chaudhury}(2013)}]{chak13}%
  \BibitemOpen
  \bibfield  {author} {\bibinfo {author} {\bibfnamefont {A.}~\bibnamefont
  {Chakrabarti}}\ and\ \bibinfo {author} {\bibfnamefont {M.~K.}\ \bibnamefont
  {Chaudhury}},\ }\href@noop {} {\bibfield  {journal} {\bibinfo  {journal}
  {Langmuir}\ }\textbf {\bibinfo {volume} {29}},\ \bibinfo {pages} {6926}
  (\bibinfo {year} {2013})}\BibitemShut {NoStop}%
\bibitem [{\citenamefont {Henann}\ and\ \citenamefont
  {Bertoldi}(2014)}]{hena14}%
  \BibitemOpen
  \bibfield  {author} {\bibinfo {author} {\bibfnamefont {D.~L.}\ \bibnamefont
  {Henann}}\ and\ \bibinfo {author} {\bibfnamefont {K.}~\bibnamefont
  {Bertoldi}},\ }\href@noop {} {\bibfield  {journal} {\bibinfo  {journal} {Soft
  Matter}\ }\textbf {\bibinfo {volume} {10}},\ \bibinfo {pages} {709} (\bibinfo
  {year} {2014})}\BibitemShut {NoStop}%
\bibitem [{\citenamefont {Style}\ and\ \citenamefont
  {Dufresne}(2012)}]{styl12}%
  \BibitemOpen
  \bibfield  {author} {\bibinfo {author} {\bibfnamefont {R.~W.}\ \bibnamefont
  {Style}}\ and\ \bibinfo {author} {\bibfnamefont {E.~R.}\ \bibnamefont
  {Dufresne}},\ }\href@noop {} {\bibfield  {journal} {\bibinfo  {journal} {Soft
  Matter}\ }\textbf {\bibinfo {volume} {8}},\ \bibinfo {pages} {7177} (\bibinfo
  {year} {2012})}\BibitemShut {NoStop}%
\bibitem [{\citenamefont {Style}\ \emph
  {et~al.}(2013{\natexlab{a}})\citenamefont {Style}, \citenamefont {Che},
  \citenamefont {Wettlaufer}, \citenamefont {Wilen},\ and\ \citenamefont
  {Dufresne}}]{styl13}%
  \BibitemOpen
  \bibfield  {author} {\bibinfo {author} {\bibfnamefont {R.~W.}\ \bibnamefont
  {Style}}, \bibinfo {author} {\bibfnamefont {Y.}~\bibnamefont {Che}}, \bibinfo
  {author} {\bibfnamefont {J.~S.}\ \bibnamefont {Wettlaufer}}, \bibinfo
  {author} {\bibfnamefont {L.~A.}\ \bibnamefont {Wilen}}, \ and\ \bibinfo
  {author} {\bibfnamefont {E.~R.}\ \bibnamefont {Dufresne}},\ }\href@noop {}
  {\bibfield  {journal} {\bibinfo  {journal} {Phys. Rev. Lett.}\ }\textbf
  {\bibinfo {volume} {110}},\ \bibinfo {pages} {066103} (\bibinfo {year}
  {2013}{\natexlab{a}})}\BibitemShut {NoStop}%
\bibitem [{\citenamefont {Style}\ \emph
  {et~al.}(2013{\natexlab{b}})\citenamefont {Style}, \citenamefont {Che},
  \citenamefont {Park}, \citenamefont {Weon}, \citenamefont {Je}, \citenamefont
  {Hyland}, \citenamefont {German}, \citenamefont {Power}, \citenamefont
  {Wilen}, \citenamefont {Wettlaufer},\ and\ \citenamefont
  {Dufresne}}]{styl13b}%
  \BibitemOpen
  \bibfield  {author} {\bibinfo {author} {\bibfnamefont {R.~W.}\ \bibnamefont
  {Style}}, \bibinfo {author} {\bibfnamefont {Y.}~\bibnamefont {Che}}, \bibinfo
  {author} {\bibfnamefont {S.~J.}\ \bibnamefont {Park}}, \bibinfo {author}
  {\bibfnamefont {B.~M.}\ \bibnamefont {Weon}}, \bibinfo {author}
  {\bibfnamefont {J.~H.}\ \bibnamefont {Je}}, \bibinfo {author} {\bibfnamefont
  {C.}~\bibnamefont {Hyland}}, \bibinfo {author} {\bibfnamefont {G.~K.}\
  \bibnamefont {German}}, \bibinfo {author} {\bibfnamefont {M.~P.}\
  \bibnamefont {Power}}, \bibinfo {author} {\bibfnamefont {L.~A.}\ \bibnamefont
  {Wilen}}, \bibinfo {author} {\bibfnamefont {J.~S.}\ \bibnamefont
  {Wettlaufer}}, \ and\ \bibinfo {author} {\bibfnamefont {E.~R.}\ \bibnamefont
  {Dufresne}},\ }\href@noop {} {\bibfield  {journal} {\bibinfo  {journal}
  {Proc. Nat. Acad. Sci.}\ }\textbf {\bibinfo {volume} {110}},\ \bibinfo
  {pages} {12541} (\bibinfo {year} {2013}{\natexlab{b}})}\BibitemShut {NoStop}%
\bibitem [{\citenamefont {Nadermann}\ \emph {et~al.}(2013)\citenamefont
  {Nadermann}, \citenamefont {Hui},\ and\ \citenamefont {Jagota}}]{nade13}%
  \BibitemOpen
  \bibfield  {author} {\bibinfo {author} {\bibfnamefont {N.}~\bibnamefont
  {Nadermann}}, \bibinfo {author} {\bibfnamefont {C.-Y.}\ \bibnamefont {Hui}},
  \ and\ \bibinfo {author} {\bibfnamefont {A.}~\bibnamefont {Jagota}},\ }\href
  {\doibase 10.1073/pnas.1304587110} {\bibfield  {journal} {\bibinfo  {journal}
  {Proc. Nat. Acad. Sci.}\ }\textbf {\bibinfo {volume} {110}},\ \bibinfo
  {pages} {10541} (\bibinfo {year} {2013})}\BibitemShut {NoStop}%
\bibitem [{\citenamefont {Bostwick}\ \emph {et~al.}(2014)\citenamefont
  {Bostwick}, \citenamefont {Shearer},\ and\ \citenamefont {Daniels}}]{bost14}%
  \BibitemOpen
  \bibfield  {author} {\bibinfo {author} {\bibfnamefont {J.~B.}\ \bibnamefont
  {Bostwick}}, \bibinfo {author} {\bibfnamefont {M.}~\bibnamefont {Shearer}}, \
  and\ \bibinfo {author} {\bibfnamefont {K.~E.}\ \bibnamefont {Daniels}},\
  }\href@noop {} {\bibfield  {journal} {\bibinfo  {journal} {Soft Matter}\ ,\ }
  (\bibinfo {year} {2014})}\BibitemShut {NoStop}%
\bibitem [{\citenamefont {Karpitschka}\ \emph {et~al.}(2014)\citenamefont
  {Karpitschka}, \citenamefont {Das}, \citenamefont {Andreotti},\ and\
  \citenamefont {Snoeijer}}]{karp14}%
  \BibitemOpen
  \bibfield  {author} {\bibinfo {author} {\bibfnamefont {S.}~\bibnamefont
  {Karpitschka}}, \bibinfo {author} {\bibfnamefont {S.}~\bibnamefont {Das}},
  \bibinfo {author} {\bibfnamefont {B.}~\bibnamefont {Andreotti}}, \ and\
  \bibinfo {author} {\bibfnamefont {J.}~\bibnamefont {Snoeijer}},\ }\href@noop
  {} {\enquote {\bibinfo {title} {{Dynamic Contact Angle of a Soft Solid}},}\ }
  (\bibinfo {year} {2014}),\ \Eprint {http://arxiv.org/abs/arXiv:1406.5547
  [physics.flu-dyn]} {arXiv:arXiv:1406.5547 [physics.flu-dyn]} \BibitemShut
  {NoStop}%
\bibitem [{\citenamefont {Johnson}\ \emph {et~al.}(1971)\citenamefont
  {Johnson}, \citenamefont {Kendall},\ and\ \citenamefont {Roberts}}]{john71}%
  \BibitemOpen
  \bibfield  {author} {\bibinfo {author} {\bibfnamefont {K.}~\bibnamefont
  {Johnson}}, \bibinfo {author} {\bibfnamefont {K.}~\bibnamefont {Kendall}}, \
  and\ \bibinfo {author} {\bibfnamefont {A.}~\bibnamefont {Roberts}},\
  }\href@noop {} {\bibfield  {journal} {\bibinfo  {journal} {Proc. Roy. Soc.
  A}\ }\textbf {\bibinfo {volume} {324}},\ \bibinfo {pages} {301} (\bibinfo
  {year} {1971})}\BibitemShut {NoStop}%
\bibitem [{\citenamefont {Style}\ \emph
  {et~al.}(2013{\natexlab{c}})\citenamefont {Style}, \citenamefont {Hyland},
  \citenamefont {Boltyanskiy}, \citenamefont {Wettlaufer},\ and\ \citenamefont
  {Dufresne}}]{styl13c}%
  \BibitemOpen
  \bibfield  {author} {\bibinfo {author} {\bibfnamefont {R.~W.}\ \bibnamefont
  {Style}}, \bibinfo {author} {\bibfnamefont {C.}~\bibnamefont {Hyland}},
  \bibinfo {author} {\bibfnamefont {R.}~\bibnamefont {Boltyanskiy}}, \bibinfo
  {author} {\bibfnamefont {J.~S.}\ \bibnamefont {Wettlaufer}}, \ and\ \bibinfo
  {author} {\bibfnamefont {E.~R.}\ \bibnamefont {Dufresne}},\ }\href@noop {}
  {\bibfield  {journal} {\bibinfo  {journal} {Nature Commun.}\ }\textbf
  {\bibinfo {volume} {4}},\ \bibinfo {pages} {2728} (\bibinfo {year}
  {2013}{\natexlab{c}})}\BibitemShut {NoStop}%
\bibitem [{\citenamefont {Salez}\ \emph {et~al.}(2013)\citenamefont {Salez},
  \citenamefont {Benzaquen},\ and\ \citenamefont {Raphael}}]{sale13}%
  \BibitemOpen
  \bibfield  {author} {\bibinfo {author} {\bibfnamefont {T.}~\bibnamefont
  {Salez}}, \bibinfo {author} {\bibfnamefont {M.}~\bibnamefont {Benzaquen}}, \
  and\ \bibinfo {author} {\bibfnamefont {E.}~\bibnamefont {Raphael}},\
  }\href@noop {} {\bibfield  {journal} {\bibinfo  {journal} {Soft Matter}\
  }\textbf {\bibinfo {volume} {9}},\ \bibinfo {pages} {10699} (\bibinfo {year}
  {2013})}\BibitemShut {NoStop}%
\bibitem [{\citenamefont {Xu}\ \emph {et~al.}(2014)\citenamefont {Xu},
  \citenamefont {Jagota},\ and\ \citenamefont {Hui}}]{xu14}%
  \BibitemOpen
  \bibfield  {author} {\bibinfo {author} {\bibfnamefont {X.}~\bibnamefont
  {Xu}}, \bibinfo {author} {\bibfnamefont {A.}~\bibnamefont {Jagota}}, \ and\
  \bibinfo {author} {\bibfnamefont {C.-Y.}\ \bibnamefont {Hui}},\ }\href@noop
  {} {\bibfield  {journal} {\bibinfo  {journal} {Soft Matter}\ }\textbf
  {\bibinfo {volume} {10}},\ \bibinfo {pages} {4625} (\bibinfo {year}
  {2014})}\BibitemShut {NoStop}%
\bibitem [{\citenamefont {Cao}\ \emph {et~al.}(2014)\citenamefont {Cao},
  \citenamefont {Stevens},\ and\ \citenamefont {Dobrynin}}]{cao14}%
  \BibitemOpen
  \bibfield  {author} {\bibinfo {author} {\bibfnamefont {Z.}~\bibnamefont
  {Cao}}, \bibinfo {author} {\bibfnamefont {M.~J.}\ \bibnamefont {Stevens}}, \
  and\ \bibinfo {author} {\bibfnamefont {A.~V.}\ \bibnamefont {Dobrynin}},\
  }\href@noop {} {\bibfield  {journal} {\bibinfo  {journal} {Macromolecules}\
  }\textbf {\bibinfo {volume} {47}},\ \bibinfo {pages} {3203} (\bibinfo {year}
  {2014})}\BibitemShut {NoStop}%
\bibitem [{\citenamefont {Style}\ \emph
  {et~al.}(2014{\natexlab{a}})\citenamefont {Style}, \citenamefont
  {Boltyanskiy}, \citenamefont {Allen}, \citenamefont {Jensen}, \citenamefont
  {Foote}, \citenamefont {Wettlaufer},\ and\ \citenamefont
  {Dufresne}}]{styl14b}%
  \BibitemOpen
  \bibfield  {author} {\bibinfo {author} {\bibfnamefont {R.~W.}\ \bibnamefont
  {Style}}, \bibinfo {author} {\bibfnamefont {R.}~\bibnamefont {Boltyanskiy}},
  \bibinfo {author} {\bibfnamefont {B.}~\bibnamefont {Allen}}, \bibinfo
  {author} {\bibfnamefont {K.~E.}\ \bibnamefont {Jensen}}, \bibinfo {author}
  {\bibfnamefont {H.~P.}\ \bibnamefont {Foote}}, \bibinfo {author}
  {\bibfnamefont {J.~S.}\ \bibnamefont {Wettlaufer}}, \ and\ \bibinfo {author}
  {\bibfnamefont {E.~R.}\ \bibnamefont {Dufresne}},\ }\href@noop {} {\bibfield
  {journal} {\bibinfo  {journal} {arXiv preprint arXiv:1407.6424}\ } (\bibinfo
  {year} {2014}{\natexlab{a}})}\BibitemShut {NoStop}%
\bibitem [{\citenamefont {Ducloue}\ \emph {et~al.}(2014)\citenamefont
  {Ducloue}, \citenamefont {Pitois}, \citenamefont {Goyon}, \citenamefont
  {Chateau},\ and\ \citenamefont {Ovarlez}}]{ducl14}%
  \BibitemOpen
  \bibfield  {author} {\bibinfo {author} {\bibfnamefont {L.}~\bibnamefont
  {Ducloue}}, \bibinfo {author} {\bibfnamefont {O.}~\bibnamefont {Pitois}},
  \bibinfo {author} {\bibfnamefont {J.}~\bibnamefont {Goyon}}, \bibinfo
  {author} {\bibfnamefont {X.}~\bibnamefont {Chateau}}, \ and\ \bibinfo
  {author} {\bibfnamefont {G.}~\bibnamefont {Ovarlez}},\ }\href@noop {}
  {\bibfield  {journal} {\bibinfo  {journal} {Soft Matter}\ }\textbf {\bibinfo
  {volume} {10}},\ \bibinfo {pages} {5093} (\bibinfo {year}
  {2014})}\BibitemShut {NoStop}%
\bibitem [{\citenamefont {Duan}\ \emph
  {et~al.}(2007{\natexlab{a}})\citenamefont {Duan}, \citenamefont {Yi},
  \citenamefont {Huang},\ and\ \citenamefont {Wang}}]{duan07a}%
  \BibitemOpen
  \bibfield  {author} {\bibinfo {author} {\bibfnamefont {H.~L.}\ \bibnamefont
  {Duan}}, \bibinfo {author} {\bibfnamefont {X.}~\bibnamefont {Yi}}, \bibinfo
  {author} {\bibfnamefont {Z.~P.}\ \bibnamefont {Huang}}, \ and\ \bibinfo
  {author} {\bibfnamefont {J.}~\bibnamefont {Wang}},\ }\href@noop {} {\bibfield
   {journal} {\bibinfo  {journal} {Mech. Mater.}\ }\textbf {\bibinfo {volume}
  {39}},\ \bibinfo {pages} {81} (\bibinfo {year}
  {2007}{\natexlab{a}})}\BibitemShut {NoStop}%
\bibitem [{\citenamefont {Duan}\ \emph
  {et~al.}(2007{\natexlab{b}})\citenamefont {Duan}, \citenamefont {Yi},
  \citenamefont {Huang},\ and\ \citenamefont {Wang}}]{duan07b}%
  \BibitemOpen
  \bibfield  {author} {\bibinfo {author} {\bibfnamefont {H.~L.}\ \bibnamefont
  {Duan}}, \bibinfo {author} {\bibfnamefont {X.}~\bibnamefont {Yi}}, \bibinfo
  {author} {\bibfnamefont {Z.~P.}\ \bibnamefont {Huang}}, \ and\ \bibinfo
  {author} {\bibfnamefont {J.}~\bibnamefont {Wang}},\ }\href@noop {} {\bibfield
   {journal} {\bibinfo  {journal} {Mech. Mater.}\ }\textbf {\bibinfo {volume}
  {39}},\ \bibinfo {pages} {94} (\bibinfo {year}
  {2007}{\natexlab{b}})}\BibitemShut {NoStop}%
\bibitem [{\citenamefont {Brisard}\ \emph
  {et~al.}(2010{\natexlab{a}})\citenamefont {Brisard}, \citenamefont
  {Dormieux},\ and\ \citenamefont {Kondo}}]{bris10}%
  \BibitemOpen
  \bibfield  {author} {\bibinfo {author} {\bibfnamefont {S.}~\bibnamefont
  {Brisard}}, \bibinfo {author} {\bibfnamefont {L.}~\bibnamefont {Dormieux}}, \
  and\ \bibinfo {author} {\bibfnamefont {D.}~\bibnamefont {Kondo}},\
  }\href@noop {} {\bibfield  {journal} {\bibinfo  {journal} {Comp. Mater.
  Sci.}\ }\textbf {\bibinfo {volume} {50}},\ \bibinfo {pages} {403} (\bibinfo
  {year} {2010}{\natexlab{a}})}\BibitemShut {NoStop}%
\bibitem [{\citenamefont {Brisard}\ \emph
  {et~al.}(2010{\natexlab{b}})\citenamefont {Brisard}, \citenamefont
  {Dormieux},\ and\ \citenamefont {Kondo}}]{bris10a}%
  \BibitemOpen
  \bibfield  {author} {\bibinfo {author} {\bibfnamefont {S.}~\bibnamefont
  {Brisard}}, \bibinfo {author} {\bibfnamefont {L.}~\bibnamefont {Dormieux}}, \
  and\ \bibinfo {author} {\bibfnamefont {D.}~\bibnamefont {Kondo}},\
  }\href@noop {} {\bibfield  {journal} {\bibinfo  {journal} {Comp. Mater.
  Sci.}\ }\textbf {\bibinfo {volume} {48}},\ \bibinfo {pages} {589} (\bibinfo
  {year} {2010}{\natexlab{b}})}\BibitemShut {NoStop}%
\bibitem [{\citenamefont {Hui}\ and\ \citenamefont {Jagota}(2013)}]{hui13}%
  \BibitemOpen
  \bibfield  {author} {\bibinfo {author} {\bibfnamefont {C.-Y.}\ \bibnamefont
  {Hui}}\ and\ \bibinfo {author} {\bibfnamefont {A.}~\bibnamefont {Jagota}},\
  }\href@noop {} {\bibfield  {journal} {\bibinfo  {journal} {Langmuir}\
  }\textbf {\bibinfo {volume} {29}},\ \bibinfo {pages} {11310} (\bibinfo {year}
  {2013})}\BibitemShut {NoStop}%
\bibitem [{\citenamefont {Duan}\ \emph
  {et~al.}(2005{\natexlab{b}})\citenamefont {Duan}, \citenamefont {Wang},
  \citenamefont {Huang},\ and\ \citenamefont {Karihaloo}}]{duan05b}%
  \BibitemOpen
  \bibfield  {author} {\bibinfo {author} {\bibfnamefont {H.}~\bibnamefont
  {Duan}}, \bibinfo {author} {\bibfnamefont {J.}~\bibnamefont {Wang}}, \bibinfo
  {author} {\bibfnamefont {Z.}~\bibnamefont {Huang}}, \ and\ \bibinfo {author}
  {\bibfnamefont {B.}~\bibnamefont {Karihaloo}},\ }\href@noop {} {\bibfield
  {journal} {\bibinfo  {journal} {J. Mech. Phys. Solids}\ }\textbf {\bibinfo
  {volume} {53}},\ \bibinfo {pages} {1574} (\bibinfo {year}
  {2005}{\natexlab{b}})}\BibitemShut {NoStop}%
\bibitem [{\citenamefont {Palierne}(1990)}]{pali90}%
  \BibitemOpen
  \bibfield  {author} {\bibinfo {author} {\bibfnamefont {J.~F.}\ \bibnamefont
  {Palierne}},\ }\href@noop {} {\bibfield  {journal} {\bibinfo  {journal}
  {Rheologica Acta}\ }\textbf {\bibinfo {volume} {29}},\ \bibinfo {pages} {204}
  (\bibinfo {year} {1990})}\BibitemShut {NoStop}%
\bibitem [{\citenamefont {Wang}\ \emph {et~al.}(2006)\citenamefont {Wang},
  \citenamefont {Duan}, \citenamefont {Huang},\ and\ \citenamefont
  {Karihaloo}}]{wang06}%
  \BibitemOpen
  \bibfield  {author} {\bibinfo {author} {\bibfnamefont {J.}~\bibnamefont
  {Wang}}, \bibinfo {author} {\bibfnamefont {H.}~\bibnamefont {Duan}}, \bibinfo
  {author} {\bibfnamefont {Z.}~\bibnamefont {Huang}}, \ and\ \bibinfo {author}
  {\bibfnamefont {B.}~\bibnamefont {Karihaloo}},\ }\href@noop {} {\bibfield
  {journal} {\bibinfo  {journal} {Proc. Roy. Soc. A}\ }\textbf {\bibinfo
  {volume} {462}},\ \bibinfo {pages} {1355} (\bibinfo {year}
  {2006})}\BibitemShut {NoStop}%
\bibitem [{\citenamefont {Style}\ \emph
  {et~al.}(2014{\natexlab{b}})\citenamefont {Style}, \citenamefont
  {Boltyanskiy}, \citenamefont {German}, \citenamefont {Hyland}, \citenamefont
  {MacMinn}, \citenamefont {Mertz}, \citenamefont {Wilen}, \citenamefont {Xu},\
  and\ \citenamefont {Dufresne}}]{styl14}%
  \BibitemOpen
  \bibfield  {author} {\bibinfo {author} {\bibfnamefont {R.~W.}\ \bibnamefont
  {Style}}, \bibinfo {author} {\bibfnamefont {R.}~\bibnamefont {Boltyanskiy}},
  \bibinfo {author} {\bibfnamefont {G.~K.}\ \bibnamefont {German}}, \bibinfo
  {author} {\bibfnamefont {C.}~\bibnamefont {Hyland}}, \bibinfo {author}
  {\bibfnamefont {C.~W.}\ \bibnamefont {MacMinn}}, \bibinfo {author}
  {\bibfnamefont {A.~F.}\ \bibnamefont {Mertz}}, \bibinfo {author}
  {\bibfnamefont {L.~A.}\ \bibnamefont {Wilen}}, \bibinfo {author}
  {\bibfnamefont {Y.}~\bibnamefont {Xu}}, \ and\ \bibinfo {author}
  {\bibfnamefont {E.~R.}\ \bibnamefont {Dufresne}},\ }\href@noop {} {\bibfield
  {journal} {\bibinfo  {journal} {Soft Matter}\ }\textbf {\bibinfo {volume}
  {10}},\ \bibinfo {pages} {4047} (\bibinfo {year}
  {2014}{\natexlab{b}})}\BibitemShut {NoStop}%
\bibitem [{\citenamefont {Landau}\ and\ \citenamefont
  {Lifshitz}(1986)}]{land86}%
  \BibitemOpen
  \bibfield  {author} {\bibinfo {author} {\bibfnamefont {L.~D.}\ \bibnamefont
  {Landau}}\ and\ \bibinfo {author} {\bibfnamefont {E.~M.}\ \bibnamefont
  {Lifshitz}},\ }\href@noop {} {\emph {\bibinfo {title} {Course of Theoretical
  Physics, Volume 7: Theory of Elasticity, Third Edition}}}\ (\bibinfo
  {publisher} {Pergamon Press},\ \bibinfo {address} {London},\ \bibinfo {year}
  {1986})\BibitemShut {NoStop}%
\bibitem [{\citenamefont {Liu}\ \emph {et~al.}(2014)\citenamefont {Liu},
  \citenamefont {Long},\ and\ \citenamefont {Hui}}]{liu14}%
  \BibitemOpen
  \bibfield  {author} {\bibinfo {author} {\bibfnamefont {T.}~\bibnamefont
  {Liu}}, \bibinfo {author} {\bibfnamefont {R.}~\bibnamefont {Long}}, \ and\
  \bibinfo {author} {\bibfnamefont {C.-Y.}\ \bibnamefont {Hui}},\ }\href@noop
  {} {\bibfield  {journal} {\bibinfo  {journal} {Soft Matter}\ ,\ \bibinfo
  {pages} {Accepted}} (\bibinfo {year} {2014})}\BibitemShut {NoStop}%
\bibitem [{\citenamefont {Schwarz}\ and\ \citenamefont
  {Safran}(2013)}]{schw13}%
  \BibitemOpen
  \bibfield  {author} {\bibinfo {author} {\bibfnamefont {U.~S.}\ \bibnamefont
  {Schwarz}}\ and\ \bibinfo {author} {\bibfnamefont {S.~A.}\ \bibnamefont
  {Safran}},\ }\href@noop {} {\bibfield  {journal} {\bibinfo  {journal} {Rev.
  Mod. Phys.}\ }\textbf {\bibinfo {volume} {85}},\ \bibinfo {pages} {1327}
  (\bibinfo {year} {2013})}\BibitemShut {NoStop}%
\bibitem [{\citenamefont {Lurie}\ and\ \citenamefont {Belyaev}(2005)}]{luri05}%
  \BibitemOpen
  \bibfield  {author} {\bibinfo {author} {\bibfnamefont {A.}~\bibnamefont
  {Lurie}}\ and\ \bibinfo {author} {\bibfnamefont {A.}~\bibnamefont
  {Belyaev}},\ }\href@noop {} {\emph {\bibinfo {title} {Theory of
  Elasticity}}}\ (\bibinfo  {publisher} {Springer, Berlin},\ \bibinfo {year}
  {2005})\ p.\ \bibinfo {pages} {246}\BibitemShut {NoStop}%
\bibitem [{\citenamefont {Milton}\ \emph {et~al.}(2006)\citenamefont {Milton},
  \citenamefont {Briane},\ and\ \citenamefont {Willis}}]{milt06}%
  \BibitemOpen
  \bibfield  {author} {\bibinfo {author} {\bibfnamefont {G.~W.}\ \bibnamefont
  {Milton}}, \bibinfo {author} {\bibfnamefont {M.}~\bibnamefont {Briane}}, \
  and\ \bibinfo {author} {\bibfnamefont {J.~R.}\ \bibnamefont {Willis}},\
  }\href@noop {} {\bibfield  {journal} {\bibinfo  {journal} {New J. Phys.}\
  }\textbf {\bibinfo {volume} {8}},\ \bibinfo {pages} {248} (\bibinfo {year}
  {2006})}\BibitemShut {NoStop}%
\bibitem [{\citenamefont {Kundu}\ and\ \citenamefont {Crosby}(2009)}]{kund09}%
  \BibitemOpen
  \bibfield  {author} {\bibinfo {author} {\bibfnamefont {S.}~\bibnamefont
  {Kundu}}\ and\ \bibinfo {author} {\bibfnamefont {A.~J.}\ \bibnamefont
  {Crosby}},\ }\href@noop {} {\bibfield  {journal} {\bibinfo  {journal} {Soft
  Matter}\ }\textbf {\bibinfo {volume} {5}},\ \bibinfo {pages} {3963} (\bibinfo
  {year} {2009})}\BibitemShut {NoStop}%
\bibitem [{\citenamefont {Cui}\ \emph {et~al.}(2011)\citenamefont {Cui},
  \citenamefont {Lee}, \citenamefont {Delbos}, \citenamefont {McManus},\ and\
  \citenamefont {Crosby}}]{cui11}%
  \BibitemOpen
  \bibfield  {author} {\bibinfo {author} {\bibfnamefont {J.}~\bibnamefont
  {Cui}}, \bibinfo {author} {\bibfnamefont {C.~H.}\ \bibnamefont {Lee}},
  \bibinfo {author} {\bibfnamefont {A.}~\bibnamefont {Delbos}}, \bibinfo
  {author} {\bibfnamefont {J.~J.}\ \bibnamefont {McManus}}, \ and\ \bibinfo
  {author} {\bibfnamefont {A.~J.}\ \bibnamefont {Crosby}},\ }\href@noop {}
  {\bibfield  {journal} {\bibinfo  {journal} {Soft Matter}\ }\textbf {\bibinfo
  {volume} {7}},\ \bibinfo {pages} {7827} (\bibinfo {year} {2011})}\BibitemShut
  {NoStop}%
\bibitem [{\citenamefont {Camp{\`a}s}\ \emph {et~al.}(2014)\citenamefont
  {Camp{\`a}s}, \citenamefont {Mammoto}, \citenamefont {Hasso}, \citenamefont
  {Sperling}, \citenamefont {O'Connell}, \citenamefont {Bischof}, \citenamefont
  {Maas}, \citenamefont {Weitz}, \citenamefont {Mahadevan},\ and\ \citenamefont
  {Ingber}}]{camp14}%
  \BibitemOpen
  \bibfield  {author} {\bibinfo {author} {\bibfnamefont {O.}~\bibnamefont
  {Camp{\`a}s}}, \bibinfo {author} {\bibfnamefont {T.}~\bibnamefont {Mammoto}},
  \bibinfo {author} {\bibfnamefont {S.}~\bibnamefont {Hasso}}, \bibinfo
  {author} {\bibfnamefont {R.~A.}\ \bibnamefont {Sperling}}, \bibinfo {author}
  {\bibfnamefont {D.}~\bibnamefont {O'Connell}}, \bibinfo {author}
  {\bibfnamefont {A.~G.}\ \bibnamefont {Bischof}}, \bibinfo {author}
  {\bibfnamefont {R.}~\bibnamefont {Maas}}, \bibinfo {author} {\bibfnamefont
  {D.~A.}\ \bibnamefont {Weitz}}, \bibinfo {author} {\bibfnamefont
  {L.}~\bibnamefont {Mahadevan}}, \ and\ \bibinfo {author} {\bibfnamefont
  {D.~E.}\ \bibnamefont {Ingber}},\ }\href@noop {} {\bibfield  {journal}
  {\bibinfo  {journal} {Nature Methods}\ }\textbf {\bibinfo {volume} {11}},\
  \bibinfo {pages} {183} (\bibinfo {year} {2014})}\BibitemShut {NoStop}%
\bibitem [{\citenamefont {Brown}\ \emph {et~al.}(1998)\citenamefont {Brown},
  \citenamefont {Prajapati}, \citenamefont {McGrouther}, \citenamefont
  {Yannas},\ and\ \citenamefont {Eastwood}}]{brow98}%
  \BibitemOpen
  \bibfield  {author} {\bibinfo {author} {\bibfnamefont {R.}~\bibnamefont
  {Brown}}, \bibinfo {author} {\bibfnamefont {R.}~\bibnamefont {Prajapati}},
  \bibinfo {author} {\bibfnamefont {D.}~\bibnamefont {McGrouther}}, \bibinfo
  {author} {\bibfnamefont {I.}~\bibnamefont {Yannas}}, \ and\ \bibinfo {author}
  {\bibfnamefont {M.}~\bibnamefont {Eastwood}},\ }\href@noop {} {\bibfield
  {journal} {\bibinfo  {journal} {J. Cell. Physiol.}\ }\textbf {\bibinfo
  {volume} {175}},\ \bibinfo {pages} {323} (\bibinfo {year}
  {1998})}\BibitemShut {NoStop}%
\bibitem [{\citenamefont {Zemel}\ \emph {et~al.}(2006)\citenamefont {Zemel},
  \citenamefont {Bischofs},\ and\ \citenamefont {Safran}}]{zeme06}%
  \BibitemOpen
  \bibfield  {author} {\bibinfo {author} {\bibfnamefont {A.}~\bibnamefont
  {Zemel}}, \bibinfo {author} {\bibfnamefont {I.}~\bibnamefont {Bischofs}}, \
  and\ \bibinfo {author} {\bibfnamefont {S.}~\bibnamefont {Safran}},\
  }\href@noop {} {\bibfield  {journal} {\bibinfo  {journal} {Phys. Rev. Lett.}\
  }\textbf {\bibinfo {volume} {97}},\ \bibinfo {pages} {128103} (\bibinfo
  {year} {2006})}\BibitemShut {NoStop}%
\bibitem [{\citenamefont {Abbena}\ \emph {et~al.}(2006)\citenamefont {Abbena},
  \citenamefont {Salamon},\ and\ \citenamefont {Gray}}]{abbe06}%
  \BibitemOpen
  \bibfield  {author} {\bibinfo {author} {\bibfnamefont {E.}~\bibnamefont
  {Abbena}}, \bibinfo {author} {\bibfnamefont {S.}~\bibnamefont {Salamon}}, \
  and\ \bibinfo {author} {\bibfnamefont {A.}~\bibnamefont {Gray}},\ }\href@noop
  {} {\emph {\bibinfo {title} {Modern Differential Geometry of Curves and
  Surfaces with Mathematica}}}\ (\bibinfo  {publisher} {Taylor \& Francis},\
  \bibinfo {year} {2006})\BibitemShut {NoStop}%
\end{thebibliography}
\end{document}